\documentclass[prc,amsmath,twocolumn,showpacs,superscriptaddress]{revtex4}
\usepackage[dvips]{graphicx}

\begin{document}

\title{$B$(E1) Strengths from Coulomb Excitation of $^{11}$Be}

\author{N.~C.~Summers}
 \affiliation{Department of Physics and Astronomy, Rutgers University, Piscataway, New Jersey 08854}
 \affiliation{National Superconducting Cyclotron Laboratory, Michigan State University, East Lansing, Michigan 48824}
\author{S.~D.~Pain}
 \affiliation{Department of Physics and Astronomy, Rutgers University, Piscataway, New Jersey 08854}
 \affiliation{Department of Physics, University of Surrey, Guildford, GU2 7XH, UK}
\author{N.~A.~Orr}
\affiliation{LPSC-Caen, ENSICAEN et Universit\'{e} de Caen, IN2P3-CNRS, 14050 Caen Cedex, France}
\author{W.~N.~Catford}
 \affiliation{Department of Physics, University of Surrey, Guildford, GU2 7XH, UK}
 \affiliation{LPSC-Caen, ENSICAEN et Universit\'{e} de Caen, IN2P3-CNRS, 14050 Caen Cedex, France}
\author{J.~C.~Ang\'elique}
 \altaffiliation{present address: LPSC-Grenoble, IN2P3-CNRS, F-38026 Grenoble Cedex, France}
 \affiliation{LPSC-Caen, ENSICAEN et Universit\'{e} de Caen, IN2P3-CNRS, 14050 Caen Cedex, France}
\author{N.~I.~Ashwood}
 \affiliation{School of Physics and Astronomy, University of Birmingham, Edgbaston, Birmingham, B15 2TT, UK}
\author{V.~Bouchat}
 \affiliation{Universit\'{e} Libre de Bruxelles, CP226, B-1050 Bruxelles, Belgium}
\author{N.~M.~Clarke}
 \affiliation{School of Physics and Astronomy, University of Birmingham, Edgbaston, Birmingham, B15 2TT, UK}
\author{N.~Curtis}
 \affiliation{School of Physics and Astronomy, University of Birmingham, Edgbaston, Birmingham, B15 2TT, UK}
\author{M.~Freer}
 \affiliation{School of Physics and Astronomy, University of Birmingham, Edgbaston, Birmingham, B15 2TT, UK}
\author{B.~R.~Fulton}
 \affiliation{Department of Physics, University of York, Heslington, York, YO10 5DD, UK}
\author{F.~Hanappe}
 \affiliation{Universit\'{e} Libre de Bruxelles, CP226, B-1050 Bruxelles, Belgium}
\author{M.~Labiche}
 \altaffiliation{present address: CCLRC Daresbury Laboratory, Daresbury, Warrington, Cheshire, WA4 4AD, UK}
 \affiliation{Electronic Engineering and Physics, University of Paisley, High Street, Paisley, Scotland, PA1 2BE, UK}
\author{J.~L.~Lecouey}
 \affiliation{LPSC-Caen, ENSICAEN et Universit\'{e} de Caen, IN2P3-CNRS, 14050 Caen Cedex, France}
\author{R.~C.~Lemmon}
 \affiliation{CCLRC Daresbury Laboratory, Daresbury, Warrington, Cheshire, WA4 4AD, UK}
\author{D.~Mahboub}
 \affiliation{Department of Physics, University of Surrey, Guildford, GU2 7XH, UK}
\author{A.~Ninane}
 \affiliation{Slashdev Integrated Solutions, B-5030 Gembloux, Belgium}
\author{G.~Normand}
 \affiliation{LPSC-Caen, ENSICAEN et Universit\'{e} de Caen, IN2P3-CNRS, 14050 Caen Cedex, France}
\author{F.~M.~Nunes}
 \affiliation{National Superconducting Cyclotron Laboratory, Michigan State University, East Lansing, Michigan 48824}
 \affiliation{Department of Physics and Astronomy, Michigan State University, East Lansing, Michigan 48824}
\author{N.~Soi\'c}
 \altaffiliation{present address: Rudjer Bo\v{s}kovi\'{c} Institute, Bijeni\v{c}ka 54, HR-10000 Zagreb, Croatia}
 \affiliation{School of Physics and Astronomy, University of Birmingham, Edgbaston, Birmingham, B15 2TT, UK}
\author{L.~Stuttge}
 \affiliation{IPHC-Strasbourg, IN2P3-CNRS et Universit\'{e} de Louis Pasteur, BP28, 67037 Strasbourg Cedex, France}
\author{C.~N.~Timis}
 \affiliation{Department of Physics, University of Surrey, Guildford, GU2 7XH, UK}
\author{I.~J.~Thompson}
 \affiliation{LLNL, P.O. Box 808, Livermore, California 94550}
 \affiliation{Department of Physics, University of Surrey, Guildford, GU2 7XH, UK}
\author{J.~S.~Winfield}
 \altaffiliation{present address: Gesellschaft f\"{u}r Schwerionenforschung GSI, D-64291 Darmstadt, Germany}
 \affiliation{Department of Physics, University of Surrey, Guildford, GU2 7XH, UK}
 \affiliation{LPSC-Caen, ENSICAEN et Universit\'{e} de Caen, IN2P3-CNRS, 14050 Caen Cedex, France}
\author{V.~Ziman}
 \affiliation{School of Physics and Astronomy, University of Birmingham, Edgbaston, Birmingham, B15 2TT, UK}

\date{\today}

\begin{abstract}
The $B$(E1;$1/2^+\to1/2^-$) strength for $^{11}$Be has been extracted from intermediate energy Coulomb excitation measurements, over a range of beam energies using a new reaction model, the extended continuum discretized coupled channels (XCDCC) method.
In addition, a measurement of the excitation cross section for $^{11}$Be+$^{208}$Pb at 38.6 MeV/nucleon is reported.
The $B$(E1) strength of 0.105(12) e$^2$fm$^2$ derived from this measurement is consistent with those made previously at 60 and 64 MeV/nucleon, in contrast to an anomalously low result obtained at 43 MeV/nucleon.
By coupling a multi-configuration description of the projectile structure with realistic reaction theory, the XCDCC model provides for the first time a fully quantum mechanical description of Coulomb excitation.
The XCDCC calculations reveal that the excitation process involves significant contributions from nuclear, continuum, and higher-order effects.
An analysis of the present and two earlier intermediate energy measurements yields a combined B(E1) strength of 0.105(7) e$^2$fm$^2$.
This value is in good agreement with the value deduced independently from the lifetime of the $1/2^-$ state in $^{11}$Be, and has a comparable precision.
\end{abstract}


\maketitle

The transition strengths between the lowest energy states of nuclei are amongst their most fundamental properties, being related to the configuration of the states, and consequently to particular aspects of the NN interaction.
As such, transition strengths are especially important for nuclei near the limits of stability, and provide important input for nuclear astrophysics.
Lifetime measurements, whenever possible, are often the preferred method for measuring transition strengths.
A less direct method of determining these strengths is Coulomb excitation, whereby the projectile is excited via the field of virtual photons produced by a heavy target \cite{glas98}.
One of the advantages of the Coulomb excitation method is that it can be applied to a wider variety of nuclei.
It is thus important to validate the method for the well known dripline cases.
Such work has been performed for quadrupole transitions on mostly stable nuclei \cite{cook06}.
In this paper we focus on dipole transitions applied to an unstable system, and resolve a long standing experimental anomaly.

The $^{11}$Be system is an archetype of a halo nucleus and exhibits the fastest known dipole transition between bound states in nuclei.
The $B$(E1) transition strength between the ground and the only bound excited state was determined from lifetime measurements by Millener {\it et al.} to be 0.116(12) e$^2$fm$^2$ \cite{millener83}.

The anomalously low value for the Coulomb excitation cross section from a lead target at 43 MeV/nucleon \cite{anne95} produced a cross section approximately 40\% of that expected from Alder and Winther theory of Coulomb excitation \cite{alderwinther}.
Subsequently, further experiments have been performed to verify this measurement, utilizing different targets and energies; MSU at $\sim$60 MeV/nucleon on Pb,Au,Be,C \cite{fauerbach97}, RIKEN at 64 MeV/nucleon on Pb \cite{nakamura97}.
The results of these measurements showed broad agreement with the lifetime measurement, although at the lower limit of the uncertainties.
However, all of the subsequent measurements have been performed at significantly ($\sim$50\%) higher energy than the anomalous GANIL measurement.

Previous analyses of Coulomb excitation experiments to extract the $B$(E1) of $^{11}$Be \cite{anne95,fauerbach97,nakamura97} have relied upon the Alder and Winther theory of Coulomb excitation \cite{alderwinther}.
A summary of $B$(E1) strengths extracted using Alder-Winther theory was given in Fig.~2 of Ref.~\cite{fauerbach97}.
Although this theory gives a useful proportional relationship between the $B$(E1) and the cross section, many approximations are applied.
The Alder and Winther approach assumes that the reaction occurs as a single-step excitation induced by the virtual photons forming the Coulomb field of the target.

There have been many theoretical efforts to investigate whether this anomaly could be explained by continuum, coupled channel \cite{bertulani95} and multi-step effects \cite{typel95}, a more accurate treatment of the nuclear absorption \cite{tarutina03}, and relativistic effects \cite{bertulani03}.
As summarized in Ref.~\cite{fauerbach97}, estimates for corrections to Alder-Winther theory at the GANIL energy were; continuum effects 4\% \cite{bertulani95}, higher order effects 6--11\% \cite{typel95} and 8\% \cite{anne95}, and nuclear effects 8\% \cite{anne95}.
None of these theoretical studies could explain the anomalously low cross section of Anne {\it et al.}~\cite{anne95}.
The theoretical uncertainty from extracting the $B$(E1) strengths using Alder-Winther theory of Coulomb excitation arising from all of these corrections is not included in the uncertainty quoted (10-13\%) \cite{anne95,fauerbach97,nakamura97}, as this relates to the experimental uncertainty only.
Additionally, a recent analysis of nuclear effects by Hussein {\it et al.}~\cite{hussein06}, shows that estimating the nuclear contribution to the Coulomb excitation cross section from the corresponding cross section on the light targets is unreliable.

The effects of these approximations have been examined individually, but a consistent analysis which includes all of them simultaneously has so far not been applied.
One main difficulty is that fully quantum mechanical reaction theories, such as the continuum discretized coupled channels (CDCC) method \cite{sakuragi86}, treat the projectile of interest as a single particle state: in the case of $^{11}$Be, a $2s_{1/2}$ neutron coupled to an inert $^{10}$Be core.
Under the constraint of a reasonable r.m.s.\ radius for the ground state of $^{11}$Be, a single particle structure model for $^{11}$Be over-estimates the $B$(E1) transition strength between the bound states by a factor of two and thus the predicted cross section is too large (see Set 0 Table~\ref{TABLE:pot}).
Therefore in order to extract a reasonable $B$(E1) strength from the cross section, one must rely on strongly model-dependent
assumptions.

Recent advances in reaction modelling \cite{summers06} allow for an improved structure input in the reaction theory.
This is the first application of the extended continuum discretized coupled channels (XCDCC) method developed in Ref.~\cite{summers06} to Coulomb excitation.
The particle-rotor structure model, as applied to $^{11}$Be in Ref.~\cite{nunes96a}, yields an accurate $B$(E1) strength which can then be consistently included in the reaction theory using the formalism developed in Ref.~\cite{summers06}.
Thus a reliable calculation for the inelastic cross sections can be made for the first time using a fully quantum mechanical description of the scattering process.

Here we report a new experiment performed at 38.6 MeV/nucleon on a lead target at GANIL.
The measurement presented here was undertaken in an attempt to resolve any experimental problems and to elucidate any possible energy dependence in the extraction of the $B$(E1).
We compare this measurement to theory predictions using XCDCC, from which we extract a $B$(E1) strength.
We also investigate the effects of various couplings within the XCDCC framework to explore deviations from the standard Coulomb excitation theory estimated in previous theoretical studies \cite{bertulani95,typel95,tarutina03,bertulani03}.
Finally, we re-analyze results from all previous experiments and compare the $B$(E1) strengths extracted with that obtained from the lifetime measurement.

\begin{figure}[t]
\includegraphics[width=8cm,angle=0]{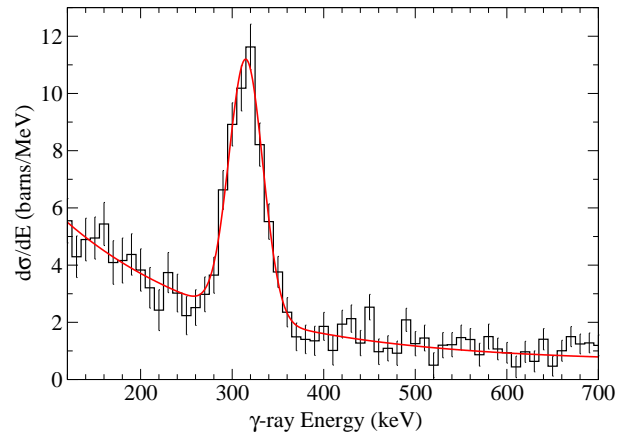}
\vspace{0.1cm}
\caption{\label{FIG:gamma} (Color online). The Doppler corrected, background subtracted $\gamma$-ray energy spectrum in coincidence with $^{11}$Be ions identified in the 0$^{\circ}$ telescope, for a $^{11}$Be beam on a Pb target at a mid-target energy of 38.6 MeV/nucleon. The 320 keV ${1/2}^-$ to ${1/2}^+$ transition in $^{11}$Be is clearly identifiable. The full line is the result of a Gaussian fit with an exponential plus constant background.}
\end{figure}

\vspace{1ex}
The data presented here were obtained utilizing a secondary beam of $^{11}$Be at GANIL, produced by the fragmentation of a primary beam of $^{18}$O at 63 MeV/nucleon on a $^9$Be production target.
The fragmentation products were purified using the LISE3 spectrometer, resulting in a beam of $\sim$5000 $^{11}$Be ions per second with 97\% purity, at a mean energy of 41 MeV/nucleon with an energy spread of $\sim$1\%.
The time-of-flight of the ions through the LISE spectrometer was measured event-by-event using a PPAC located at the entrance to the reaction chamber together with an MCP, $\sim$23 m upstream, and also relative to the cyclotron RF.
These time-of-flight measurements allowed the unique selection of $^{11}$Be ions event-by-event from the 3\% impurities in the beam (mainly $^9$Li and $^{13}$B).

The $^{11}$Be beam was incident on a 552 mg/cm$^2$ Pb target, resulting in a mid-target beam energy of 38.6 MeV/nucleon.
Charged particles were detected in a three-stage telescope centered at 0$^{\circ}$, consisting of two 500 $\mu$m thick resistive-strip silicon detectors (providing two $\Delta$E measurements, and measurement of position in vertical ($x$) and horizontal ($y$) directions), backed by a closely packed array of 16 CsI detectors in a 4$\times$4 arrangement \cite{Ahmed}.
The telescope subtended the angular range $\pm 9^\circ $ in $x$ and $y$.
Gamma rays were detected, in coincidence with charged particles in the telescope, using four NaI detectors at $\pm 45^\circ$ and $\pm 110^\circ$ to the beam axis, giving a total absolute photopeak efficiency of 3.1\% for detection of Doppler-shifted 320 keV $\gamma$-rays (assuming an isotropic distribution in the center-of-mass system and including relativistic forward-focusing effects of order 18\% and absorption in the target of order 2\%).
Background data were taken with no target present in order to determine the contribution from reactions occurring in the telescope itself.
These were normalized to, and subtracted from, the target-in data.
The present data was acquired contiguously with another experiment utilizing the same setup \cite{pain06}; the same calibration and analysis procedures were used for the present measurement. 

The Doppler corrected $\gamma$-ray energy spectrum in coincidence with $^{11}$Be, after background subtraction, is shown in Fig.~\ref{FIG:gamma} --- a peak corresponding to the ${1/2}^-$ to ${1/2}^+$ transition ($E_\gamma = 320$keV) is clearly evident, yielding a cross section of 416(44) mb. The quoted error reflects uncertainties in the fitting procedure, the detector efficiencies, the integrated number of $^{11}$Be beam particles and the target thickness. Next we compare with reaction theory predictions in order to extract the $B$(E1).

\begin{table}\begin{tabular}{c|cccccc|cc}
      & $V_{\mathrm{even}}$ & $V_{\mathrm{odd}}$   & $V_{so}$ & $R$ & $a$ & $\beta_2$ & $B$(E1) & $\sigma$ \\ \hline
0   & 55.07    & 36.25 & 6.4   & 2.736 & 0.67  & 0    & 0.282 & 1092\\
I   & 54.24    & 49.67 & 8.5   & 2.483 & 0.65  & 0.67 & 0.139 & 549 \\
II  & 55.22    & 50.82 & 8.5   & 2.483 & 0.60  & 0.67 & 0.125 & 493 \\
III & 55.04    & 48.10 & 5.0   & 2.483 & 0.65  & 0.67 & 0.116 & 459 \\
IV  & 55.70    & 48.86 & 5.0   & 2.483 & 0.60  & 0.67 & 0.106 & 421 \\
V   & 56.29    & 49.51 & 5.0   & 2.483 & 0.55  & 0.67 & 0.096 & 383 \\ 
\end{tabular}
\caption{\label{TABLE:pot}
$^{10}$Be-n Woods-Saxon potential parameters with energies in MeV and lengths in fm, along with the calculated $B$(E1) strength in e$^2$fm$^2$, and inelastic cross section in mb.
Set 0 is the Be12-pure and Set I is the Be12-b, both from Ref.~\cite{nunes96a}. The experimental cross section is 416(44) mb.}
\end{table}

An accurate prediction of the $B$(E1) strength can be obtained in the particle-rotor model, where $^{11}$Be is treated as a deformed $^{10}$Be coupled to a neutron (see Set I Table~\ref{TABLE:pot}) \cite{nunes96a}.
This two-body structure model assumes $^{10}$Be is a rotor and includes its $2^+$ excited state explicitly.
It predicts around 15\% excited core contributions in the ground state of $^{11}$Be, in good agreement with no-core shell model calculations \cite{forssen05}, and experiment \cite{aumann00,fortier99,winfield01,palit03}. 
It also produces $B$(E1)=0.139 e$^2$fm$^2$ between the two bound states, which
is close to the value extracted from the lifetime measurement.

The $^{11}$Be description we use in the XCDCC calculations is the model described above.
We consider a set of parameters based on Be12-b from Table I of Ref.~\cite{nunes96a}, which are given in Set I of Table~\ref{TABLE:pot}.
Later in the paper we will consider the sensitivity of the cross sections to various parameterizations of this potential which yield varying $B$(E1) strengths.
The neutron-target interaction used is the global optical potential of Schandt {\it et al.}~\cite{schandt82}.
The $^{10}$Be-Pb central interaction is taken from Ref.~\cite{fukuda04}.
In XCDCC calculations, this  interaction is further deformed using $\beta_2^n=0.67$ for the nuclear part and $\beta_2^C=1.13$ for the Coulomb, consistent with Ref.~\cite{nunes96b}.
When re-analyzing all previous Coulomb excitation data, we use the same $^{10}$Be-target
potential parameters but scale the radius parameters from the Pb to the Au target.
Errors arising from uncertainties in the nuclear optical potentials are small \cite{capel03}.

Next we give a brief description of the model space for the XCDCC calculation.
For the $^{11}$Be bound and continuum states, we take all partial waves up to $\ell_{\mathrm{max}}=3$.
The $^{10}$Be-n continuum is discretized up to ${J^{\pi}_P}_{\mathrm{max}}=5/2^+$ and $E_{\mathrm{rel}}\le5$ MeV.
As we are only interested in the inelastic cross section to the bound excited state, the details of the continuum are not important and thus it is sufficient to use rather broad energy bins.
For this reason 40 bins are included, 5 bins for $J_P^\pi=3/2^-,5/2^+$, with ground state core components in the boundary conditions, and 2 bins for all other partial waves.
Radial integrals for the continuum bins were performed out to 50 fm with a step length of 0.01 fm.

\begin{table}\begin{tabular}{l|c|c}
      & \multicolumn{2}{c}{$\sigma_{\mathrm{Coulex}}$ (mb)} \\ \cline{2-3}
beam energy (MeV/nucleon) & 39 & 59  \\ \hline
Pure Coulomb one-step                    & 706 & 500 \\
+ nuclear absorption                     & 579 & 421 \\
+ nuclear in the excitation coupling     & 687 & 499 \\
+ bound states coupled to all orders     & 642 & 469 \\
+ continuum coupled to all orders        & 506 & 407 \\
+ dynamical core excitation (full XCDCC) & 549 & 415 \\
\end{tabular}
\caption{\label{TABLE:calc}
Calculated $^{11}$Be+Pb Coulomb excitation cross sections at 39 MeV/nucleon (this work) and at 59 MeV/nucleon (MSU) showing the effects of the progressive improvements in the reaction calculation (see text).}
\end{table}

In Table~\ref{TABLE:calc} we present a summary of the various types of reaction calculation that can be performed for this system taking the $^{11}$Be structure model directly from \cite{nunes96a}.
By turning off couplings within XCDCC, we can provide an indication of the size of the various effects which have previously been estimated. 

We start with the quantum formulation closest to the semi-classical Alder-Winther theory: pure Coulomb one-step, taking only Coulomb interactions for incoming and outgoing waves, as well as for the transition operator.
This would be similar to the DWBA formulation with Coulomb waves, except that here we go beyond Alder-Winther theory as the projectile is described within a coupled channels model.
The cross section of 706 mb (Table~\ref{TABLE:calc}) obtained with this simplistic approach is unrealistically large.
This is well understood as arising from the lack of nuclear absorption: in the Alder-Winther theory a lower limit on the impact parameter is introduced to account for this.
The sensitivity of the cross section on this is investigated in Ref.~\cite{tarutina03}, where a cutoff obtained from the particle-rotor structure model is applied.
When introducing nuclear absorption into the XCDCC calculation by including a diagonal nuclear potential in the $^{11}$Be+Pb optical potential, the cross section is reduced to 579 mb.

The nuclear potential in the distorted waves has a negligible effect on the cross section.
By subsequently including the nuclear interaction in the transition operator, the cross section is increased by 19\%.
Note that nuclear excitation cross section to the $1/2^-$ state yields 27 mb.
This reaffirms that nuclear effects cannot be simply subtracted, as nuclear-Coulomb interference is very large \cite{hussein06}.
Coupling the two bound states to all orders reduces the cross section by 7\%.
A reduction of 21\% is obtained with the inclusion of the continuum, resulting in a cross section of 506 mb.
Finally, XCDCC can include the dynamical excitation of the core during the reaction, which has not been studied previously.
This coupling allows the $^{10}$Be core to change spin through its interaction with the target.
The ground and excited states of $^{11}$Be have different amounts of excited core contributions and therefore this coupling allows the relative amounts of excited core contributions to re-adjust, enhancing the cross section by 9\%.
Including this effect, which corresponds to the full XCDCC calculation, yields a final cross section of 549 mb.

It is remarkable that the full calculation produces a cross section similar (within 5\%) to the Coulomb-only one-step calculation with nuclear distortion (second line in Table \ref{TABLE:calc}).
This is due to the strong increase arising from the nuclear interaction being (coincidentally in the current case) nearly canceled by the inclusion of the continuum.

The importance of nuclear effects has little dependence on the beam energy and precise scattering angles.
An analysis of the MSU experiment on the Pb target reveals a similar progression of corrections (+19\%,-6\%,-13\%,+2\% for an overall change of -2\%).
The presence of effects of the order of 20\% suggests again that Alder-Winther theory should not be relied upon for the extraction of the $B$(E1).

The calculations presented here are integrated over the full angular range.
While the MSU measurement had the smallest angular acceptance, this still contained 99\% of the total inelastic cross section.
Owing to the significant increase in computation time required to get this level of accuracy on the angular distribution, the effect of the angular cut was neglected for these calculations.
For all of the other measurements the angular acceptance was larger and had a negligible effect on the inelastic cross section.

Relativistic effects are not included in the present work.
Relativistic calculations including retardation have been performed for $^{11}$Be at 50 and 100 MeV/nucleon \cite{bertulani03}.
Within a semi-classical reaction model, deviations of non-relativistic theory from a full relativistic treatment were calculated as only 2\% at 50 MeV/nucleon, and increasing up to 14\% at 100 MeV/nucleon.
Although a full relativistic quantum mechanical formulation is still lacking, for the measurement reported in this paper we therefore expect that errors arising from the non-relativistic treatment of the reaction are less than 2\%, and around 5\% for the MSU and RIKEN measurements respectively.

The full XCDCC calculations for $^{11}$Be Coulomb excitation at an energy of 38.6 MeV/nucleon 
on a lead target, using set I from Table~\ref{TABLE:pot} for n-$^{10}$Be,
produced $\sigma=549$ mb, somewhat larger than the measured cross section of 416(44) mb. 
As the structure model \cite{nunes96a} somewhat over-predicts the $B$(E1) strength (0.139 e$^2$fm$^2$ compared to  0.116(12) e$^2$fm$^2$ from lifetime measurements) we explore small variations in the parameterization.
By varying the diffuseness and spin-orbit potential depth, then adjusting the odd and even potential depths to reproduce the ground and excited state binding energies, a range of $B$(E1) strengths were obtained (sets I--V of Table~\ref{TABLE:pot}).
We note that Set III is the parameterization used in Ref.~\cite{tarutina03}.
We found a good linear relationship between the $B$(E1) strength and the Coulomb excitation cross section, as would be the case if Alder-Winther theory were valid.
The deviation from the linear regression is very small.
Finally, we find that the experimental cross section is reproduced when the model for $^{11}$Be predicts $B$(E1)=0.105(12) e$^2$fm$^2$, in good agreement with the lifetime measurement.

\begin{figure}[t]
\includegraphics[width=7.5cm]{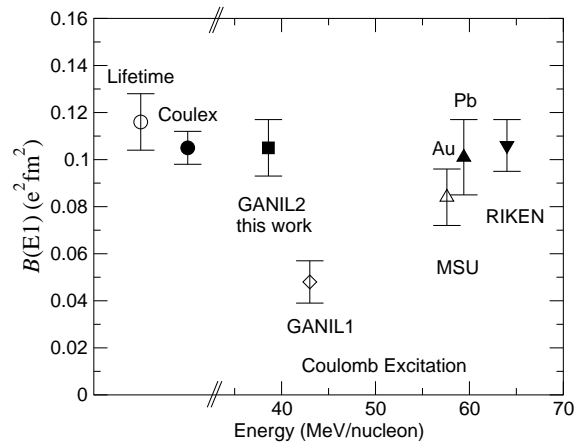}
\caption{\label{FIG:be1-expt} Comparison of $B$(E1) values obtained from lifetime and Coulomb excitation measurements. The weighted average of lifetime measurements \cite{millener83} (open circle) is plotted on the left along with the weighted average (solid cicle) of three Coulomb excitation measurements (solid symbols). The individial Coulomb excitation measurements, GANIL (this work, square), MSU (up triangle) \cite{fauerbach97}, RIKEN (down triangle) \cite{nakamura97}, and a previous GANIL experiment (diamond) \cite{anne95}, are plotted versus the beam energy.}
\end{figure}

We now re-analyze the cross sections from the previous experiments and extract $B$(E1) consistently, using the same reaction model, namely XCDCC.
The results are summarized in Fig.~\ref{FIG:be1-expt}.
Our re-analysis of the previous GANIL measurement yields the much lower $B$(E1) value, 0.048(9) e$^2$fm$^2$.
The authors of this work had mentioned in their previous article ``this result remains difficult to interpret, and even if there is no indication of a possible error in the analysis, we feel that it should be regarded with some caution until it has been confirmed in a new experiment'' \cite{anne95}.
It has become apparent, based on subsequent observations of the electronic counting/dividing modules employed, that this measurement may have been affected by a problem in the normalization of the beam intensity \cite{laurent07}.

For the MSU measurement, we deduce $B$(E1)=0.101(16) e$^2$fm$^2$ and 0.084(12) e$^2$fm$^2$ for the Pb and Au targets respectively. For the RIKEN measurement, $B$(E1)= 0.106(11) e$^2$fm$^2$.
All of these are consistent with the results from the lifetime measurement.
It may be remarked that the Au target measurement, in contrast with the three Pb measurements, involved a correction of order 50\% for $\gamma$-ray absorption in the target. The three Pb results, with quoted errors of 10--15\%, are internally consistent at the 5\% level.
Taking into account the possible experimental problems with the original GANIL measurement and also the rather large target correction for the only Au data point, the remaining three measurements --- all with a Pb target --- have been combined to give a weighted average of $B$(E1)=0.105(7) e$^2$fm$^2$.

Prior to this work, one could speculate that the anomalous measurement indicated an energy dependence in the Coulomb excitation method for the extraction of the $B$(E1).
Our new measurement removes this possibility.
In summary, the work described here demonstrates the validity of the Coulomb excitation method for extracting $B$(E1) of exotic nuclei, using a fully quantum mechanical treatment of the reaction process.
Moreover, we have shown that the accuracy of $B$(E1) extracted from Coulomb excitation can be comparable with that from lifetime measurements.

This work was supported in part by NNSA through DOE Cooperative Agreement DEFC03-03NA00143, NSCL, Michigan State University, and the National Science Foundation through grant PHY-0456656, and under the auspices of the University of California, Lawrence Livermore National Laboratory under contract No.\ W-7405-Eng-48.
The authors wish to acknowledge the support provided by the technical staff of LPC and GANIL.
Partial support through the EU Human Mobility programme of the European Community is also acknowledged.
The calculations were performed at the HPCC at MSU.


\end{document}